\documentstyle[12pt,epsf]{article}

\begin{document}

\title{Heisenberg Spin Glass on a Hypercubic Cell}
\author{Daniel A. STARIOLO\\[0.5em] 
{\small Dipartimento di Fisica, Universit\`a di Roma I} 
{\small {\em ``La Sapienza''}}\\
{\small and}\\
{\small Istituto Nazionale di Fisica Nucleare, Sezione di Roma I}\\
{\small Piazzale Aldo Moro 2, 00185 Rome, Italy}\\[0.3em]
{\small \tt stariolo@chimera.roma1.infn.it}\\[0.5em] }

\date{\today}
\maketitle

\begin{abstract}
We present results of a Monte Carlo simulation of an Heisenberg Spin Glass
model on a hipercubic cell of size 2 in {\it D} dimensions. Each spin 
interacts with {\it D} nearest neighbors and the lattice is expected to 
recover the 
completely connected (mean field) limit as $D\rightarrow \infty$. An analysis 
of the Binder parameter for $D=8, 9$ and $10$ shows clear evidence of the 
presence of a spin glass phase at low temperatures. We found that in the high 
temperature regime the inverse spin glass susceptibility  grows linearly with
$T^2$ as in the mean field case. Estimates of $T_c$ from the high temperature
data are in very good agreement with the results of a Bethe-Peierls 
approximation for an Heisenberg Spin Glass with coordination number {\it D}.
\end{abstract}
\thispagestyle{empty}
\newpage

\section{\bf Introduction}
Despite the fact that many real spin glasses, like ${\rm Eu_x Sr_{1-x}S, CuMn,
 AgMn}$
, are Heisenberg like systems, the very existence of a finite temperature
transition in three dimensions seems to be ruled out for systems
with short range interactions \cite{olive}. Simulations of a model with
long range RKKY interactions are compatible with the system being at its lower
critical dimension \cite{reger}. A possible
way out of the puzzle has been the consideration of anisotropy. Matsubara et 
al. \cite{matsubara1} have shown that even a small amount of anisotropy in the
Hamiltonian is enough for having a spin glass phase at low temperatures. The
previous models \cite{olive,reger,matsubara1} are of the class with {\it bond 
disorder}. A 
finite temperature transition was also found in an isotropic {\it site 
diluted} model
with RKKY interactions \cite{matsubara2}. More recently, 
Coluzzi \cite{barbara1}
have found evidence for a finite temperature transition in the four
dimensional bond disordered lattice with nearest neighbour interactions (nni).
 Consequently 
the lower critical dimension for the model with nni would be between three and
four.

In this letter we address the problem of the spin glass transition in 
Heisenberg systems putting enphasis in the connectivy structure of the lattice
rather than the dimension. In fact, the coordination number in the real
three dimensional space may not be six for an amorphous system, as forced by 
the hypercubic lattice geometry. Instead,
we have studied a model in which $N$ spins are placed in the vertices of a 
{\it D} dimensional {\it 
hipercubic cell} of side 2 so that the size of the system is $N=2^D$ and where
each spin interacts with its {\it D} nearest neighbors. This model has been
introduced by Parisi {\it et al.} \cite{parisi} who studied the static 
properties of
Ising spin glasses and the approach to mean field behaviour that is expected
when $D \rightarrow \infty$. Note that in this geometry {\it D} is not the 
dimension
of real space but defines the connectivity structure of the system. From the
point of view of the connectivity the behaviour of the {\em hipercubic cell} 
in dimension {\it D} has to be compared with that of the {\em hipercubic 
lattice} in {\it D/2} dimensions.
For an Heisenberg system the spins interact through the Hamiltonian:
\begin{equation}
{\cal{H}}= - \frac{1}{2} \sum_{<ij>}^N \, J_{ij} \,\, {\bf s_i\cdot s_j}
\end{equation}
The vector spins $\{{\bf s_i}, i=1\ldots N\}$ have components 
$\{s_i^\alpha, 
\alpha=1\ldots 3\}$ and are normalized in the unit sphere. The random 
interactions are chosen from a Gaussian distribution with $\overline{J_{ij}}=0$
and $\overline{J_{ij}^2}= 1/D$. It is important to note that
in this geometry the size $N$ and the ``dimension'' $D$ are constrained so it 
is not possible to do the usual finite size scaling analysis by fixing $D$ and 
let $N$ grow to the thermodynamic limit. A real phase transition can only occur
in the limit $D\rightarrow \infty$ which corresponds to mean field. 
Nevertheless clear evidences of a phase transition can already be seen at 
finite size $N$ (or equivalently, finite ``dimension'' $D$). We will
 see that the present model is very
well suited for studying the approach to mean field behaviour and the effects
of finite connectivity.  

\section{\bf The Binder Parameter and Spin Glass Susceptibility}

In vector models the usual spin glass order parameter becomes a matrix in
the spin components. It is possible to define the set of overlaps
\begin{equation}
q^{\alpha \beta}= \frac{1}{N}\sum_{i=1}^N \, s_i^{\alpha}s_i^{\beta}\,,
\end{equation}
where for the Heisenberg case $\alpha=1\ldots 3$ denotes spin components and
$i=1\ldots N$ the sites on the lattice. As the system presents a global
rotational symmetry it is useful to define a rotationally invariant order
parameter Q whose moments are
\begin{equation}
Q^k=\left[ \sqrt{\frac{1}{3}\sum_{\alpha \beta}(q^{\alpha\beta})^2}\,.
\right]^k
\end{equation}
A useful quantity for determining the existence of a phase transition is the
so called Binder parameter defined, for the Heisenberg system \cite{barbara1},
as
\begin{equation}
g=\frac{1}{2}\left[11-9\,\frac{\overline{<Q^4>}}{\left(\overline{<Q^2>}
\right)^2} \right]\,,
\end{equation}
where $<\ldots>$ means a thermal average and the overbar an average over
disorder realizations of the bonds.
From the scaling properties of this adimensional quantity it turns out that
at the critical point the value of {\it g} is independent of the system size
and consequently the curves for different sizes must cross each other at $T_c$.
This fact permits a rather accurate determination of the critical temperature.
Considering two replicas of the system ${\bf \sigma}$ and ${\bf \tau}$
 it can be defined the spin glass susceptibility as
\begin{equation}
\chi_{SG}=\beta^2\frac{3}{N}\overline{\left<\left(\sum_{i=1}^N
{\bf \sigma_i \cdot \tau_i}\right)^2\right>}\,.
\end{equation}
The factor 3 has been introduced in order to have $\chi_{SG}/\beta^2=1$
for $T\rightarrow\infty$ as in the SK model \cite{fischer}. With our definition
of the parameter {\it Q} the spin glass susceptibility can be expressed
\begin{equation}
\chi_{SG}=3\,N\,\beta^2 \,\overline{< Q^2 >}\,.
\end{equation}
The spin glass susceptibility is expected to diverge at and below the spin 
glass transition temperature $T_c$.

We have done Monte Carlo simulations of the Heisenberg spin glass previously
defined and measured the moments $Q^2$ and $Q^4$ and also the Binder parameter
and spin glass susceptibility. The dynamics used has been a standard heat bath
algorithm \cite{olive}. As we are here interested in static properties we have
checked that the system was thermalized before measuring physical quantities by
looking at the coincidence of the susceptibilities calculated by two different
methods, replicas and auto-overlaps as in \cite{bhatt}. We have simulated 
systems of $N=256,\, 512$ and $1024$ spins corresponding to $D=8,\, 9$ and $10$
respectively.

\section{\bf Results}
At high thermal noise $T\rightarrow\infty$ the spins become independent
variables and the limiting values of the moments $<Q^2>$ and $<Q^4>$ can be 
determined exactly \cite{barbara2}. It is found that 
\begin{equation}
\lim_{T\to\infty}\langle Q^2\rangle = 1/3N \,,
\end{equation}
 and 
\begin{equation}
\lim_{T\to\infty}\langle Q^4 \rangle = \frac{11-2/N}{81N^2}\,.
\end{equation}
Figure 1 shows a plot of $\log \langle Q^k \rangle$ vs
$\log N=D\log 2$ for $T=5$. The solid lines show the exact predicted behaviour.
 The agreement between both results, simulations and analytic, is
very impressive given credit to both our data and the analytic result. As at
this temperature the spins are uncorrelated it is not necessary to perform the
average over disorder samples. At lower temperatures we have averaged over 50
to 100 samples.


In Figure 2 we can see the spin glass susceptibility rescaled with $T^2$
vs temperature in the range $0.15 \leq T \leq 0.5$. The behaviour suggests a
finite temperature phase transition in this range of temperatures. From
Eqs.(6) and (7) it can be seen that the curves must go to 1 as the temperature
increases.   Further
evidence is obtained from the Binder parameter shown in Figure 3. Normally the
transition temperature is evidenced by the point where the curves for
different sizes intersect. Our results are consistent with a transition at
$T_c\approx 0.3$. But the behaviour is quite peculiar; in the high T regime
$g\approx 0$ for the three sizes studied. A possible explanation for this fact
may be that the geometry of the lattice is not fixed and also the use of free
boundary conditions \cite{binder}. In this lattice all spins are also in the
boundary, so using free boundary conditions make the spins less constrained
than, for example, using periodic boundary conditions. We also recall that in 
this model $N$ and $D$ grow
together, it is not possible to fix $D$ and let $N$ grow to the thermodynamic 
limit. In this sense, the three sizes studied correspond also to three 
different ``dimensions'' of the cell, or coordination values $D$. The effect
of increasing $D$ is to move the corresponding curve slightly to the right,
and this is why we do not see a clear crossing of the succesive $D$ curves. 



A sensible
analysis can be made of the approach to the thermodynamic limit, {\it i.e.} the
infinite dimension or
completely connected model, where mean field is exact. In the simulations we
expect to see corrections to mean field behaviour coming from the finite
connectivity $D$. It is known that the 
Heisenberg spin glass in the mean field limit presents a transition at a 
critical temperature $T_c=1/3$ (for spins normalized in the unit sphere) below
which there is a continuous breaking of the replica symmetry \cite{almeida}. 
Our
results from the Binder parameter suggest a slightly lower value that can be
explained in part as a finite $D$ correction. In fact, a Bethe-Peierls
approximation for the Heisenberg spin glass with coordination $D$ predicts a
tranisition at a critical temperature given by the equation \cite{olive}:
\begin{equation}
(D-1)\,\,\overline{ \left[\coth(J_{ij}/T_c) - T_c/J_{ij}\right]^2} = 1
\end{equation}
In Table 1 we show the solutions of this equation for different $D$ values. 
Note that the results will depend strongly on the normalization chosen for the
$J_{ij}'s$, for example if we had chosen $\overline{J_{ij}^2}=1$ as usual for
short range models, the $T_c$ predicted for $D=8$ would be $T_c=0.6825$ instead
of $T_c=0.2413$, a considerable difference. 
As can be seen the $T_c$ predicted for the range of $D$ studied in our 
simulations are considerably lower than the mean field value 0.333 which 
shows that the approach to the mean field limit is slow.

\begin{table}
\centering
\begin{tabular}{|c|c|} \hline
$D$      & $T_c$       \\ \hline \hline
 6      & 0.2117   \\ \hline
 7      & 0.2288   \\ \hline
 8      & 0.2413   \\ \hline
 9      & 0.2514   \\ \hline
 10     & 0.2594   \\ \hline
 100    & 0.3259   \\ \hline
\end{tabular}
\caption{The critical temperatures predicted by the 
Bethe-Peierls approximation}
\end{table}   

The mean field spin glass susceptibility  for an Heisenberg system in the
paramagnetic phase can be shown to be:
\begin{equation}
\chi_{SG}= \frac{\beta^2}{1-(\beta/3)^2}= \frac{1}{T^2-1/9}
\end{equation}
so that the inverse susceptibility is expected to behave linearly with $T^2$.
We show in Figure 4 a plot of $\chi_{SG}^{-1}$ vs $T^2$ for the mean field
result together with the results of our simulations in the range of 
temperatures 0.35 - 0.5. The solid lines are linear fits to the data points. 
In this range a perfect linear behaviour is observed
in agreement with mean field but with slope slightly different from one.
If we consider values of $T < 0.35$ the behaviour is no longer linear
suggesting that we are departing from the high $T$ regime. 


From the linear
fits we were able to estimate by extrapolation the critical temperatures for
the different $D$ studied and compare with the results of the Bethe-Peierls
approximation. The results are summarized in Table 2 and show a very good
agreement with the analytic ones of Table 1.
\begin{table}
\centering
\begin{tabular}{|c|c|} \hline
$D$             & $T_c$       \\ \hline \hline
8             & 0.2576   \\ \hline
9             & 0.2641   \\ \hline
10            & 0.2730   \\ \hline
mean field    & 0.3333   \\ \hline
\end{tabular}
\caption{The critical temperatures from a linear fit of the high
temperature data}
\end{table}   

\section{\bf Conclusions}
We have presented evidence that isotropic Heisenberg spin glasses with finite 
connectivities present, at low temperatures, a spin glass transition for not 
too small connectivities.
The overal behaviour of the system is in qualitative agreement with mean field
predictions. At high temperatures finite size effects seem to be very weak and
the main corrections come from the finite connectivity $D$. The critical 
temperatures for $D=8, \, 9$ and $10$
depart considerably from the mean field ($D=\infty$) result. An improved 
estimation with a
Bethe-Peierls approximation, which takes into account finite connectivity
effects, is in very good agreement with the numerical results from 
extrapolations of the high temperatura data. 

We think that the hipercubic cell 
can be a useful model for studying the robustness of mean field predictions 
in the
more realistic case of short range interactions, while keeping the possibility
(specially from a computational point of view) of going to considerbly larger 
values of the connectivity than those which can be attained in ordinary 
hipercubic lattices. For the Heisenberg model it would be of
particular interest the calculation of the distribution of overlaps
$P(Q)$ which gives information of the structure of phase space and whose non
trivial character for short range interactions is still in debate. 

I would like to acknowledge useful discussions with G. Parisi, 
J. J. Ruiz-Lorenzo and B. Coluzzi.

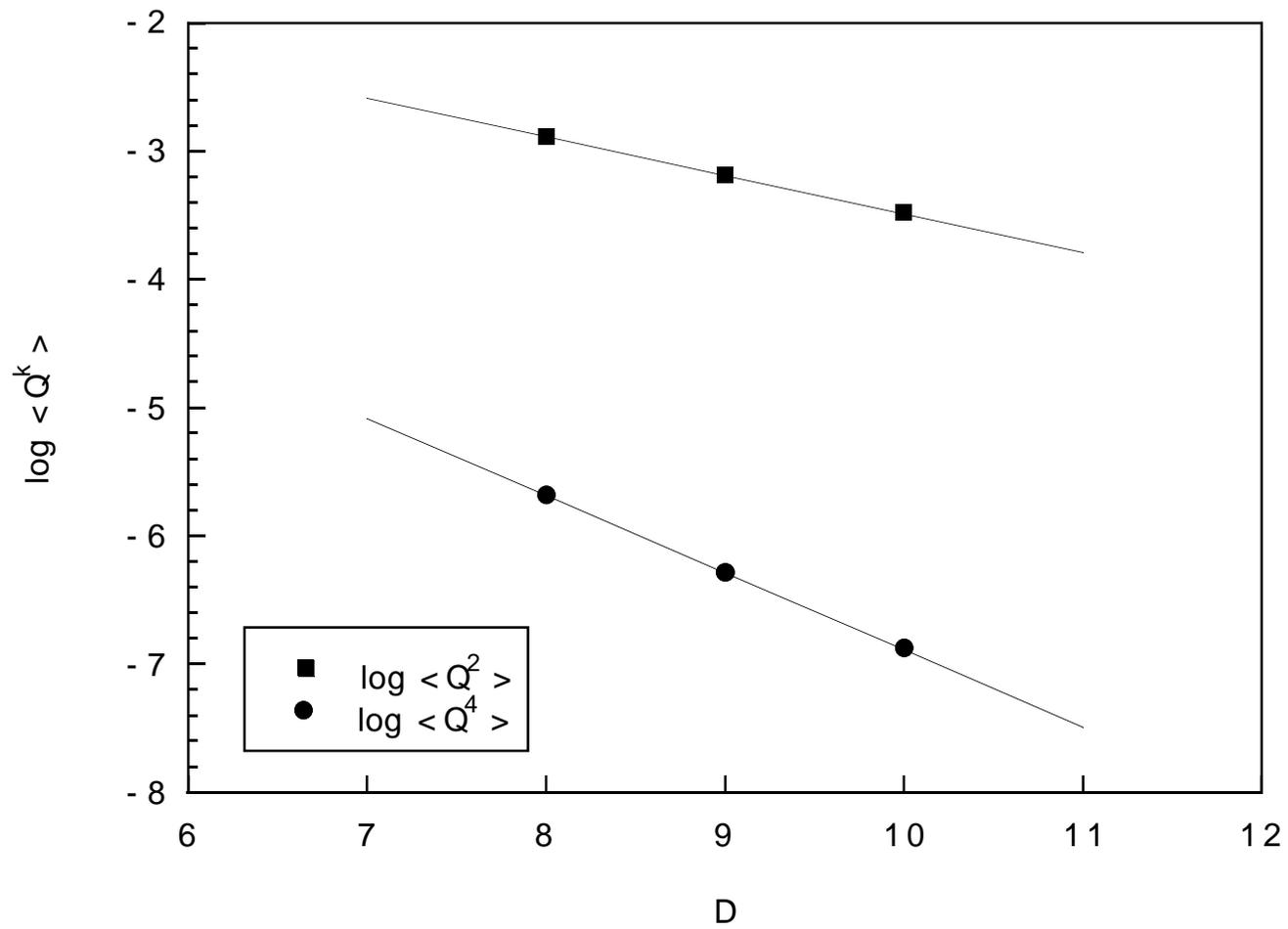
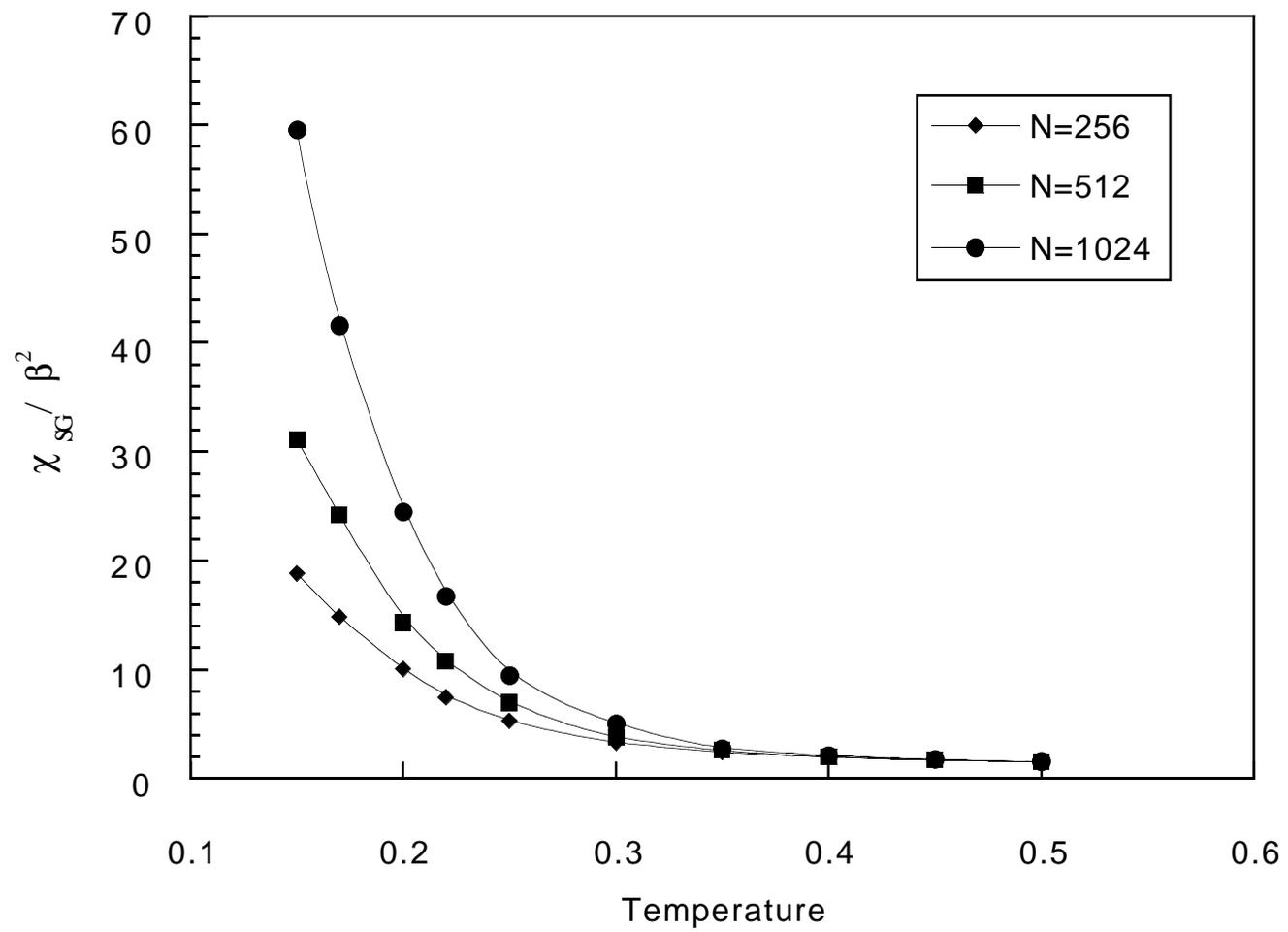
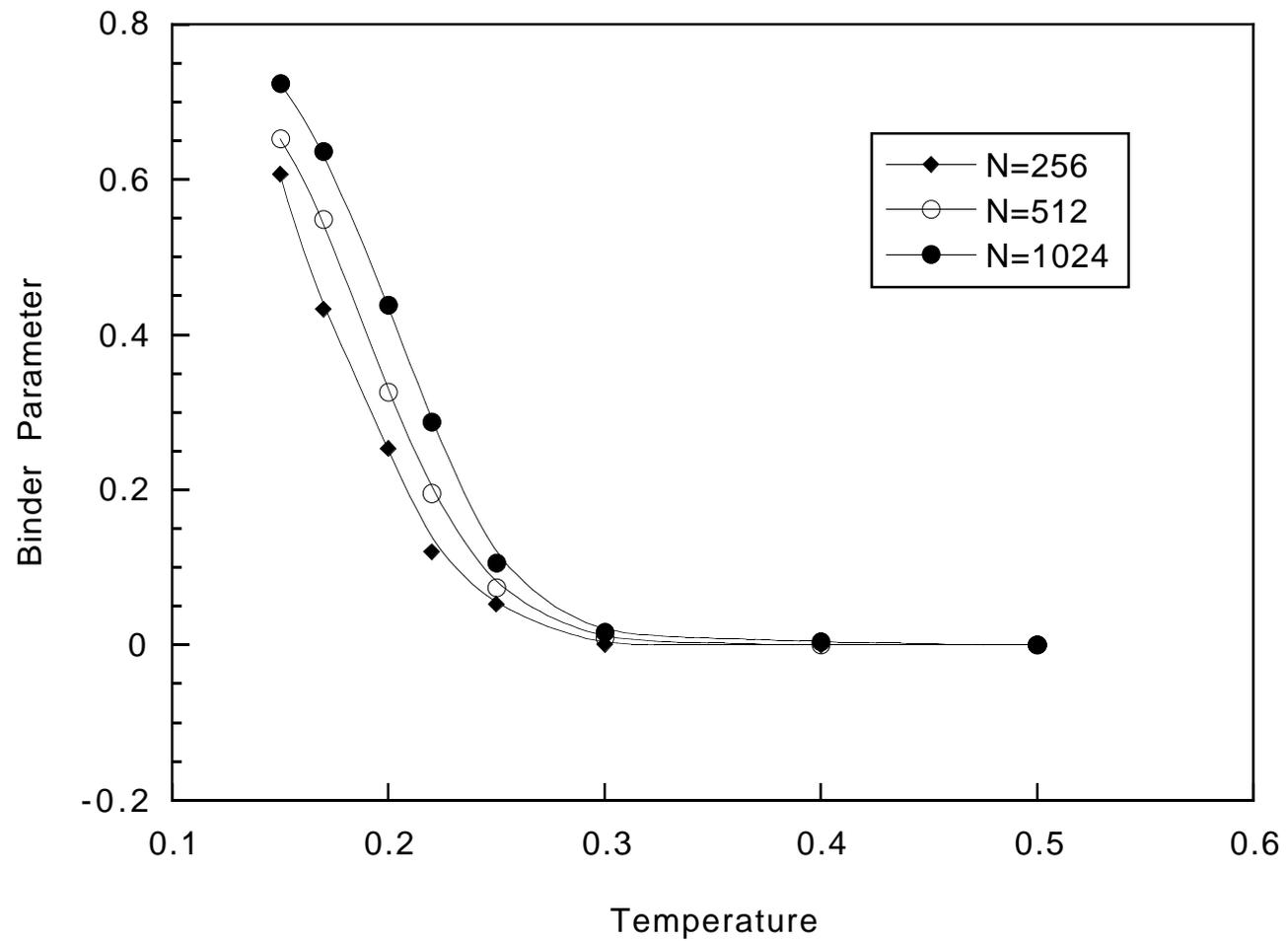
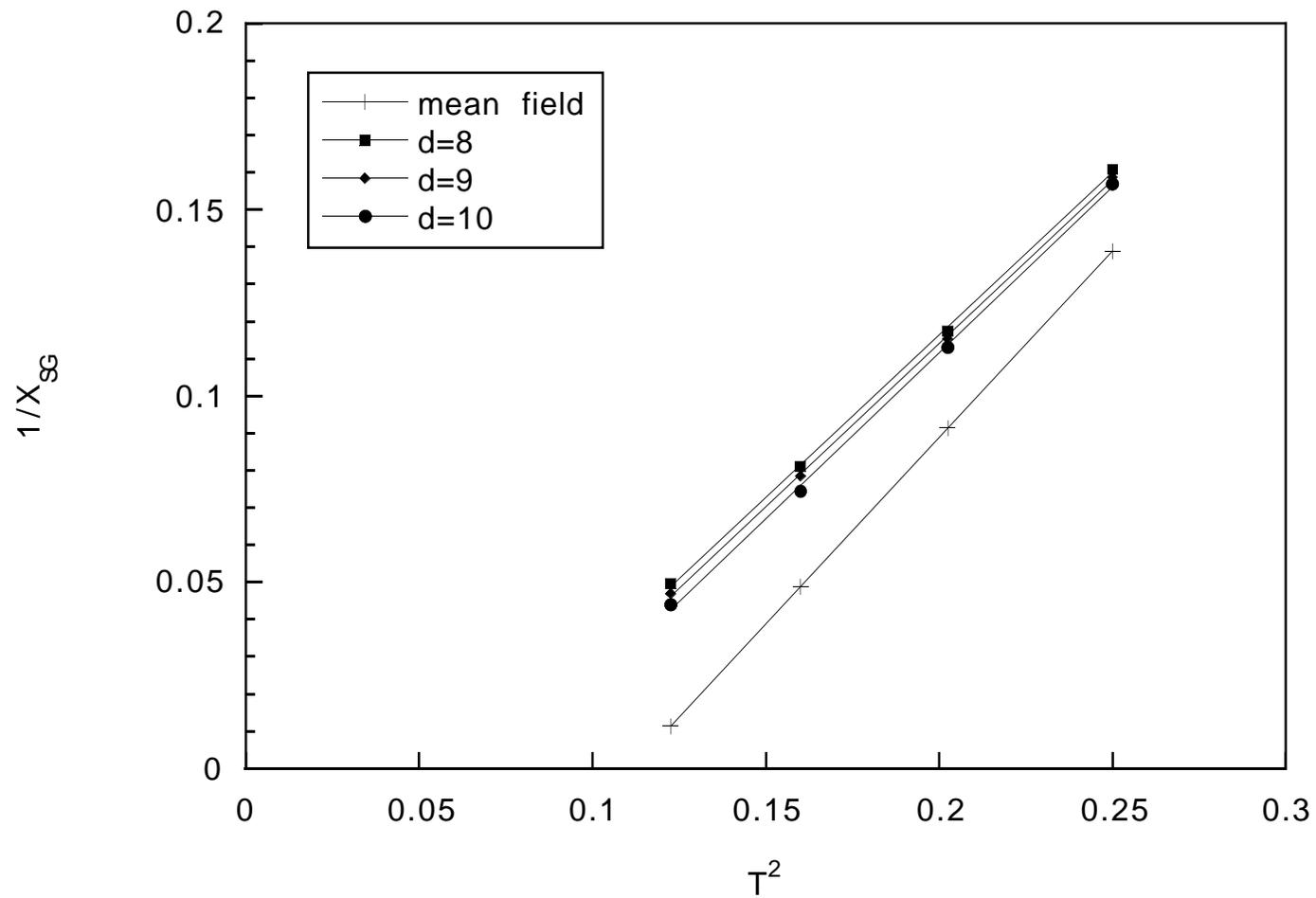
\end{document}